\documentclass[letter,twocolumn,showpacs,preprintnumbers,pra,byrevtex]{revtex4}

\usepackage{graphics}
\usepackage{amsmath}
\usepackage{amssymb}
\usepackage{bm}
\usepackage{graphicx}

\begin{document}
\title{Purity Swapping in the Jaynes-Cummings Model: Obtaining Perfect Interference Patterns from Totally Unpolarized Qubits}
\author{J. Mart\'{\i}nez-Manso and J. Mart\'{\i}nez-Linares}
\email{jesusml@us.es}
\address{Departamento de F\'{i}sica Aplicada II.Universidad de Sevilla.  \\
41012-Seville, Spain. }


\begin{abstract}
We show the existence of dynamical purity swapping phenomena in the
Jaynes-Cummings model. Moreover we show that purity swapping between
a qubit and a generic quantum system is possible, provided they are
coupled via non-unitary matrix elements interaction. We
particularize to the case of a quantum optical Ramsey
interferometer. We show that using purity swapping, a perfect
interference pattern can be obtained at the output port of the
interferometer even if we start from totally unpolarized sources.
This feature is shown to be associated with the phenomena of
recreation of the state vector at half of the revival time. In fact,
we show that the Gea-Banacloche attractor is robust against
degradation of the purity of the qubit input state. We also show
that the Tsallis entropy $T_2$ is a useful entanglement monotone
allowing one to relate directly entanglement with purity exchange in
interacting systems. We conjecture an Araki-Lieb type inequality for
$T_2$ that bounds the maximum interchange of purity between interacting systems.
\end{abstract}

\pacs{03.67.-a 03.67.Mn 07.60.Ly 42.50.Md}



\narrowtext

\maketitle

%
%
%

\section{Introduction}

The Jaynes-Cummings Model (JCM) is a paradigmatic description for
many problems involving the interaction of spin-like two-level
systems with single mode bosonic systems \cite{shore}. Examples can
be found in a large variety of systems, ranging from quantum dots
coupled to optical or microwave fields \cite{wilson} to circuit-QED,
e.g., in a Cooper-pair box of a superconducting quantum interference
device (SQUID) \cite{blais}. Another well known example is cavity
QED \cite{berman}, covering systems like Rydberg atoms in microwave
cavities \cite{Hagley97} or trapped ions cavity QED
\cite{guth,mundt}. The inherent ability of the strong coupling
regime of cavity-QED to coherently convert quantum states between
material qubits and photon qubits opens the door to a large number
of applications to quantum information processing (QIP)
\cite{raimond}.

Entanglement is a fundamental quantum non-local resource in QIP
which lacks, however, of a complete quantification \cite{chuang}. In
this paper we will show that linear entropies $\mathcal{G}$ are a
useful entanglement measure \cite{gallis96} that allows us to relate
entanglement with the the interchange of purity between interacting
systems. We conjecture \cite{footnote4} that the linear entropies
satisfy the Araki-Lieb type inequality

\begin{equation}
|\mathcal{G_A}-\mathcal{G_B}|\leq \mathcal{G_{AB}} \leq
\mathcal{G_A}+\mathcal{G_B} \label{O13.7},
\end{equation}
for a composite system $\mathcal{AB}$. Eq. (\ref{O13.7}) is an
important relation, since it bounds the possibility of a mutual
transfer of purity between interacting $\mathcal{A}$ and
$\mathcal{B}$ subsystems.

The aim of this paper is two fold. First, to study $\mathcal{G}$ as
an entanglement measure for the JCM. This will lead us to a
remarkable result: the existence of dynamical purity swapping in
this system bounded by Eq. (\ref{O13.7}). We accompany this results
with numerical simulations supporting the validity of Eq.
(\ref{O13.7}) for the JCM. Second, to find a necessary condition for
purity swapping in a general interaction between a qubit and a
generic quantum system. To achieve this goal we will follow a recent
interferometric approach \cite{Jesus04, Jesus07}, developed to keep
track of which-way information (WWI) in duality experiments. This
will allow us to answer the question: can we obtain a perfect
interference pattern starting from a totally unpolarized source?

This paper is organized as follows. In Section II we will describe
the Tsallis entropies as entanglement monotones. In Section III we
study an interferometer coupled to unitary which-way markers (WWM).
This will allow us to setup the notation and show that for the
question posed in the previous paragraph the answer is negative. In
section IV we show that this is no longer the case for non-unitary
WWM. In Section V we particularize the formalism to a QORI. We end
up with conclusions and a summary of the results.

\section{Tsallis entropies as entanglement monotones}

The entanglement of the pure states of a bipartite
$\mathcal{A},\,\mathcal{B}$ system is completely quantified by a
unique measure \cite{popescu} but only in a specific asymptotic
limit. This measure is the entropy of entanglement \cite{bennett}
\begin{equation}
E=S(\rho_\mathcal{A})=S(\rho_\mathcal{B}), \label{O0.1}
\end{equation}
where $S(\rho)=-\textup{tr}\rho\log_2\rho$ is the Von Neumann
entropy and
$\rho_{\mathcal{A},\mathcal{B}}=\textup{tr}_{\mathcal{B},\mathcal{A}}\,\rho_{\mathcal{A
B}}$ are the reduced density matrices of the
$\mathcal{A}(\mathcal{B})$ subsystem obtained after partial tracing
the overall state over the other $\mathcal{B}(\mathcal{A})$
subsystem. The entropy of entanglement satisfies the Araki-Lieb
inequality \cite{araki-lieb}
\begin{equation}
|\mathcal{S_A}-\mathcal{S_B}|\leq \mathcal{S_{AB}} \leq
\mathcal{S_A}+\mathcal{S_B} \label{O13.61}.
\end{equation}

Outside the asymptotic limit, or for mixed states, $E$ is no longer
a good measure of entanglement. Here, there are a number of
different measures of entanglement that have been proposed
\cite{bennett2}. One of them are the entanglement monotones ($EM$)
\cite{vidal2000}, which consist in any function of the quantum state
non-increasing under LOCC (local operations and classical
communication). An example of $EM$ are the $\alpha$-entropies
\cite{wehrl}
\begin{equation}
S_\alpha=\frac{1}{1-\alpha}\log_2\textup{tr}\rho^\alpha,\;\;\;\alpha\,\epsilon\,[0,1].
\label{O0.2}
\end{equation}
It is also easy to show that the non-additive Tsallis q-entropies
\cite{tsallis88}
\begin{equation}
T_q=\frac{1}{q-1}\left(1-\textup{tr}\rho^q\right) \label{O0.3}
\end{equation}
are also $EM$ measures for $q>0$. Note that the entropy of
entanglement $E$ is recovered in the limit $\alpha\rightarrow1$ and
$q\rightarrow1$, respectively. Tsallis $EM$ will prove an useful
tool for analyzing the entanglement properties of our system.
Concretely we will use $T_2$, which has a direct physical meaning,
since it is directly related to the purity of the state
$P=\textup{tr}\rho^2$. $T_2$ has been called the linear entropy
\cite{gallis96}
\begin{equation}
\mathcal{G}_\mathcal{O}=1-P_\mathcal{O}, \label{O0.4}
\end{equation}
where the subscript
$\mathcal{O}=(\mathcal{A},\,\mathcal{B},\,\mathcal{AB})$ refers to
the system under consideration. $\mathcal{G}_\mathcal{AB}$ satisfies
the non-extensive property

\begin{eqnarray}
\mathcal{G_{AB}}=\mathcal{G_A}+\mathcal{G_B}-\mathcal{G_A}\mathcal{G_B},\label{O13.3}
\end{eqnarray}
for the case of uncorrelated $\mathcal{A},\, \mathcal{B}$ subsystems
\cite{tsallis88, raggio}. Inserting Eq. (\ref{O0.4}) into
(\ref{O13.3}), this property reduces to the factorization condition
\begin{equation}
P_{\mathcal{AB}}=P_{\mathcal{A}} P_{\mathcal{B}}\label{O13.31}
\end{equation}
for the purity of a factorizable state
$\rho_{\mathcal{AB}}=\rho_{\mathcal{A}}\otimes \rho_{\mathcal{B}}$.
As a matter of fact, additivity is not an {\it a-priori} requirement
for a good measure of entanglement \cite{vidal99}. Thus, we will use
the linear entropy $\mathcal{G}=T_2$ as an $EM$. This will allow us
to relate entanglement with purity exchange in interacting systems.

\section{Two-way interferometers with unitary WWM}

Let's consider the two-way interferometer showed in Fig. 1(a).
Following \cite{Jesus04}, we describe the quanton degree of freedom
as a two-level system. Its initial state is prepared as

\begin{equation}
\rho_Q^{(0)} = \frac{1}{2} \left( 1 + \boldsymbol{s}_Q^{(0)} \cdot
\boldsymbol{\sigma} \right),  \label{1}
\end{equation}
where $\boldsymbol{\sigma} = (\sigma_x,\sigma_y,\sigma_z) $ are the
usual Pauli spin operators and $\boldsymbol{s}_Q^{(0)} =
(s_{Qx}^{(0)}, s_{Qy}^{(0)}, s_{Qz}^{(0)}) $ is the Bloch vector of
the quanton describing its initial polarization state. The norm of
the Bloch vector comprises particle-like and wave-like information.
In fact \cite{Englert96}
\begin{equation}
{|\boldsymbol{s}^{(0)}_Q|}^2
={s_{Qx}^{(0)}}^2+\left({s_{Qy}^{(0)}}^2+{s_{Qz}^{(0)}}^2\right)=\mathcal{P}^2+\mathcal{V}_{0}^2={|\boldsymbol{s}^{(f)}_Q|}^2,
\label{O2}
\end{equation}
where $\mathcal{V}_0$ is the visibility of the interference pattern
at the output port of the interferometer and $\mathcal{P}=|\omega_+
- \omega_-|=|{s}^{(0)}_{Qx}|$ \cite{footnote1} is the predictability
of the alternative ways taken by the quanton. Here $\omega_\pm$ are
the probabilities for the quanton taking the up or down ways after
passage of the beam splitter. The norm of the Bloch vector is
directly related to the purity of the state
\begin{equation}
P_Q=\text{tr}\, \rho_Q^{2}=\frac{1}{2}(1+|\boldsymbol{s}_Q|^2)
\label{O2.1},
\end{equation}
so $|\boldsymbol{s}_Q|$ is conserved at all times under unitary
evolution, i.e., $|\boldsymbol{s}^{(0)}_Q|=|\boldsymbol{s}^{(f)}_Q|$
where $f$ stands for the final state of the quanton. This is no
longer the case if a which-way-marker (WWM) is additionally inserted
in order to acquire extra which-way information (WWI), in the guise
of Fig. 1(b). Once the quanton passes through the WWM, it transforms
the marker's state as

\begin{equation}
\rho _{M}^{(0)}\rightarrow U_{\pm }^{\dag } \, \rho _{M}^{(0)} \,
U_{\pm }\equiv \rho _{M}^{(\pm )},  \label{O3}
\end{equation}
where $\rho _{M}^{(0)}$ is the initial state of the marker. $U_{+}$
and $U_{-}$ are unitary operators describing the action of the WWM.
The fringe visibility is now given \cite{Englert96} by the
expression

\begin{equation}
\mathcal{V} = |\mathcal{C}| \mathcal{V}_0 \, \label{O4},
\end{equation}
where

\begin{equation}
\mathcal{C}\equiv \text{tr}_{M}\left\{ U_{+}^{\dag }\, \rho
_{M}^{(0)} \, U_{-}\right\} \label{O5}
\end{equation}
is a contrast factor, $ 0\le |\mathcal{C}|\le 1$. Thus, the
visibility with WWM is always equal or lesser than $\mathcal{V}_0$.
This implies a degradation of the norm of the Bloch vector, which is
now given by \cite{Englert96}

\begin{equation}
{|\boldsymbol{s}^{(f)}_Q|}^2 =\mathcal{P}^2+\mathcal{V}^2,
\label{O6}
\end{equation}
Combining Eq. (\ref{O0.4}) with Eqs. (\ref{O2.1}) and (\ref{O6}) we
obtain
\begin{equation}
\Delta P_Q=P^{(f)}_Q-
P^{(0)}_Q=\frac{1}{2}\mathcal{V}^2_0\left(|\mathcal{C}|^2-1\right).\label{O6.2}
\end{equation}
Thus, the purity of the quanton always decreases or stays equal as a
result of the interaction with unitary WWM. According to Eq.
(\ref{O6.2}), the purity is conserved ($|\mathcal{C}|=1$) in the
absence of entanglement between quanton and WWM ($|U_+|=|U_-|$).
Since the purity of the quanton never increases, starting with a
totally unpolarized source ($|\boldsymbol{s}^{(0)}_Q|=0$) it is just
impossible to obtain an interference pattern in any two-way
interferometer coupled to any unitary WWM. This can be seen
explicitly in Eq. (\ref{O4}), since
$\mathcal{V}_0=\sqrt{|\boldsymbol{s}^{(0)}_{Qy}|^2+|\boldsymbol{s}^{(0)}_{Qz}|^2}=0$
in this case.
\begin{figure}[h]
\includegraphics[scale=0.4]{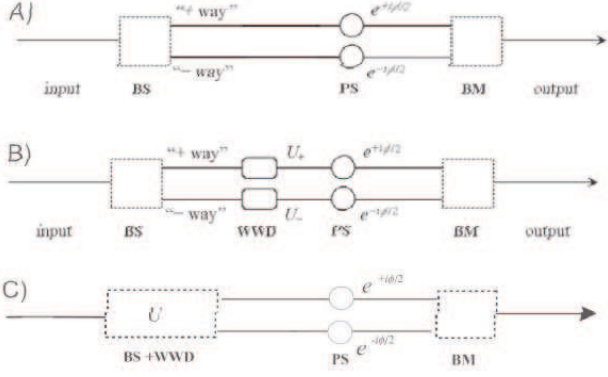} \caption{\label{fig2}  Schematic two-way interferometer setup,
composed by: (a) A Beam splitter (BS), a  Phase Shifter (PS) and a
Beam Merger (BM). (b) The interferometer is supplemented with an
additional quantum degree of freedom, the WWM, characterized by
unitary $U_{\pm}$ evolution. (c) The unitary condition of $U_{\pm}$
is released to account for more general coupling to a WWM. Here, the
evolution operator $U$ may exhibit non-unitary matrix elements.}
\end{figure}

\section{Two-way interferometers with non-unitary WWM}

Let us prepare the quanton initially in the state
$s^{(0)}_Q=(0,0,s)$, where $-1\leq s \leq 1$ is the inversion
\cite{footnote2}. Consider now the case plotted in Fig. 1(c). The
state is given initially by
$\rho^{(0)}=\rho_Q^{(0)}\otimes\rho_M^{(0)}$. The system evolves in
time, $\rho^{(0)}\longrightarrow\rho^{(f)}=U^\dagger\rho^{(0)}U$
according to the unitary operator
\begin{equation}
\label{O7}
 U = \frac{1}{\sqrt{2}} \left(
\begin{array}{cc}
V_{++} & V_{+-} \\
-V_{-+} & V_{--}
\end{array}
 \right),
\end{equation}\\
where we have followed the notation given in \cite{Englert96b}.
Although $U$ is unitary, it might not be the case for its matrix
elements separately. The particular case $V_{++}=V_{-+}=U_+,\;
V_{+-}=V_{--}=U_-$ recovers the situation described in Eq.
(\ref{O3}).

The final state of the quanton, after application of all the
transformation representing all the elements of the interferometer
given in Fig. 1(c), is calculated in \cite{Jesus07} to be

\begin{equation}
\label{O19} \rho^{(f)}=\frac{1+s}{2} \; \rho^{(+)} +\frac{1-s}{2} \;
\rho^{(-)} ,
\end{equation}

with

\begin{eqnarray}
\label{O20}
\rho_+^{(f)}=\frac{1+\sigma_x}{4}V_{++}^{\dag}\rho_M^{(0)}V_{++}+\frac{1-\sigma_x}{4}V_{+-}^{\dag}\rho_M^{(0)}V_{+-}
\nonumber\\
-\frac{\sigma_z
-i\sigma_y}{4}e^{-i\phi}V_{++}^{\dag}\rho_M^{(0)}V_{+-}-\frac{\sigma_z
+i\sigma_y}{4}e^{i\phi}V_{+-}^{\dag}\rho_M^{(0)}V_{++}, \nonumber\\
\end{eqnarray}
\\
and $\rho_-^{(f)}$ is obtained from $\rho_+^{(f)}$ through the
changes $V_{++}\rightarrow -V_{-+},\;V_{+-}\rightarrow V_{--}$.
Tracing over the cavity field degree of freedom, the final Bloch
vector of the quanton can be calculated to be

\begin{eqnarray}
S_{Qx}^{(f)} &=& w_+ -w_- \, ,
\nonumber \\
S_{Qy}^{(f)} &=& \Re e\left[\mathcal{C}e^{-i\phi }\right] ,
\nonumber\\
S_{Qz}^{(f)} &=&  i\Im m \left[\mathcal{C} e^{-i\phi}\right]
,\label{O7.1}
\end{eqnarray}
where $\phi$ is the phase induced by the phase shifter. The contrast
factor reads
\begin{equation}
 \mathcal{C} = \frac{1+s}{2} \; \mathcal{C}_{\uparrow} + \frac{1-s}{2}\;
 \mathcal{C}_{\downarrow} ,
 \label{O7.5}
\end{equation}

where

\begin{eqnarray}
 \mathcal{C}_{\uparrow}      &\equiv& i\;\mbox{tr}_D
              \, \left\{ V_{++}^{\dag }\rho_{D}^{(0)}V_{+-}\right\}= i\left< V_{+-} V_{++}^{\dag }
              \right>_0 ,
\nonumber \\
 \mathcal{C}_{\downarrow}     &\equiv&  -i\;\mbox{tr}_D
              \, \left\{ V_{-+}^{\dag }\rho_{D}^{(0)}V_{--}\right\}= -i\left< V_{--} V_{-+}^{\dag } \right>_0 .
 \label{O7.8}
\end{eqnarray}

The general form of $\mathcal{P}$ and $\mathcal{V}$ has also been
calculated in \cite{Jesus07}. They read

\begin{equation}
\mathcal{P}=|\omega_+ - \omega_-|=|s^{(f)}_{Qx}|, \label{O7.2}
\end{equation}

with

\begin{eqnarray}
w_+ &=&\frac{1+s}{4} \;
                \left< V_{++} V_{++}^{\dag } \right>_0 +\frac{1-s}{4} \;
                \left< V_{-+} V_{-+}^{\dag } \right>_0 ,
\nonumber\\
w_- &=&\frac{1+s}{4} \;
                \left< V_{+-} V_{+-}^{\dag } \right>_0 +\frac{1-s}{4} \;
                \left< V_{--} V_{--}^{\dag } \right>_0 .
 \label{O7.3}
\end{eqnarray}

and
\begin{eqnarray}
\mathcal{V} = |\mathcal{C}| \leq1. \label{O7.4}
\end{eqnarray}

Note that $\mathcal{V}_0$ does not factorize now in the right hand
side of Eq. (\ref{O7.4}) as it did in Eq. (\ref{O4}). This fact
opens the door to he possibility of obtaining an interference
pattern from an unpolarized source ($\mathcal{V}_0=0$) that will be
explored in the next section. Combining Eqs. (\ref{O7.1}) and
(\ref{O7.4}) we have
\begin{equation}
\mathcal{V}^2=|s^{(f)}_{Qy}|^2+|s^{(f)}_{Qz}|^2. \label{O7.9}
\end{equation}
Summing Eqs. (\ref{O7.9}) and (\ref{O7.2}) we get

\begin{equation}
|s^{(f)}_{Q}|^2=\mathcal{V}^2 + \mathcal{P}^2. \label{O7.10}
\end{equation}
We find that Eq. (\ref{O7.10}) is a general result, valid even in
the case of non-unitary WWM. The final state of the WWM can be
calculated after tracing $\rho^{(f)}$ over the quanton's degree of
freedom. It reads \cite{Jesus07}

\begin{equation}
\label{O7.11} \rho_M^{(f)}= \text{tr}_Q\,\rho^{(f)} =\omega_+ \;
\rho_M^{(+)} +\omega_- \; \rho^{(-)}_M ,
\end{equation}
where

\begin{eqnarray}
w_+ \rho_M^{(+)}&=& \text{tr}_Q \left\{\frac{1+\sigma_x}{2} \;
\rho^{(f)}\right\}
\nonumber\\
&=& \frac{1+s}{4}\;
                 V_{++}^{\dag} \rho_D^o V_{++}  +\frac{1-s}{4} \;
                 V_{-+}^{\dag} \rho_D^o V_{-+}   ,
\nonumber\\
w_- \rho_M^{(-)}&=& \text{tr}_Q \left\{ \frac{1-\sigma_x}{2} \;
\rho^{(f)} \right\}
\nonumber\\
&=& \frac{1+s}{4} \;
                 V_{+-}^{\dag} \rho_D^o V_{+-}  +\frac{1-s}{4} \;
                 V_{--}^{\dag} \rho_D^o V_{--}
 \label{O7.12}
\end{eqnarray}
are the contributions associated to each way.

In order to analyze the exchange of entropy between quanton and WWM
we make use of the $EM$ measures introduced in Eq. (\ref{O0.4}). The
purity of the final WWM state can be calculated with the help of
Eqs. (\ref{O7.11}) and (\ref{O7.12}) in the form

\begin{eqnarray}
P_M=\text{tr}\rho_M^{2}=\frac{(1+s)^2}{16}\;\left[\left< V_{++}
V_{++}^{\dag } \right>_0^2 +\left< V_{+-} V_{+-}^{\dag } \right>_0^2
\right.
\nonumber\\
\left. + 2\left< V_{++} V_{+-}^{\dag } \right>_0\left< V_{+-}
V_{++}^{\dag } \right>_0\right]
\nonumber\\
+ \frac{(1-s)^2}{16}\;\left[\left< V_{-+} V_{-+}^{\dag } \right>_0^2
+\left< V_{--} V_{--}^{\dag } \right>_0^2\right.
\nonumber\\
\left. + 2\left< V_{-+} V_{--}^{\dag } \right>_0\left< V_{--}
V_{-+}^{\dag } \right>_0\right]
\nonumber\\
+ \frac{(1-s^2)}{16}\;\left[2\left< V_{-+} V_{++}^{\dag }
\right>_0\left< V_{++} V_{-+}^{\dag } \right>_0\right.
\nonumber\\
\left. + 2\left< V_{--} V_{+-}^{\dag } \right>_0\left< V_{+-}
V_{--}^{\dag } \right>_0\right.
\nonumber\\
\left. + 2\left< V_{-+} V_{+-}^{\dag } \right>_0\left< V_{+-}
V_{-+}^{\dag } \right>_0\right.
\nonumber\\
\left. + 2\left< V_{++} V_{--}^{\dag } \right>_0\left< V_{--}
V_{++}^{\dag } \right>_0\right] \label{O15}
\end{eqnarray}

For the quanton's purity the calculation is much easier.
Combining Eqs. (\ref{O2.1}) and (\ref{O7.10})
, we
have

\begin{equation}
P_Q=\frac{1}{2}(1+\mathcal{P}^2+\mathcal{V}^2). \label{O16}
\end{equation}

\section{The Quantum Optical Ramsey interferometer}

We particularize now the formalism described in the previous section
to the case of a quantum optical Ramsey interferometer (QORI). This
system has been extensively studied in the literature, both
theoretically \cite{Englert96b, JesusJulio04, Scully91} and
experimentally \cite{newHaroche2001, Maitre97, Rauschenbeutel99}.
The interaction hamiltonian given by the standard Jaynes-Cummings
model (JCM) \cite{jcm}

\begin{equation}
\mathcal{H}=\hbar\Omega\left(\sigma_+a+\sigma_-a^\dagger\right),
\label{O16.1}
\end{equation}
where $\sigma_+=|e\rangle\langle g|$ and $\sigma_-=|g\rangle\langle
e|$ are the ladder operators for a two level atomic system composed
by an excited $|e\rangle$ and a ground $|g\rangle$ state. These
operators interact with a coupling strength $\Omega$ (the Rabi
frequency of the atomic transition) with a microwave cavity field
mode described by the bosonic $a,\,a^{\dagger}$ annihilation and
creation operators. Thus, a low loss cavity resonator acts jointly
as a which-way marker (WWM) and a beam splitter (BS) [see
Fig.~1(c)]. Before entering the cavity, the atom is prepared, say,
in the upper level $ |e\rangle$ (case $s=1$). The atom interacts
resonantly with the cavity field, adding a photon to its quantized
cavity mode if a transition to the lower level $|g\rangle$ occurs.
Due to the low-loss factor of the resonator, the cavity field can
keep track of the way taken by the atom since it can store for long
times the energy quantum liberated in the atomic transition
\cite{kuhr}. Thus, the same interaction both splits the beam and
makes the two ``ways'' distinguishable. Next is the turn of the
phase shifter (PS)---in the guise, for example, of an external pulse
of electric field applied at the central stage of the
interferometer.
Finally, a classical microwave field at the port of the
interferometer supplies the beam merger (BM), effecting a $\pi/2$
pulse after resonant interaction with the atom. The final state of
the atom is measured by means of state-selective field ionization
techniques at the output port of the interferometer. By varying the
phase~$\phi$ in successive repetitions of the experiment, a fringe
pattern can be built up in the detected probability for the atom to
wind up in one state or the other.\\

The evolution operators of Eq. (\ref{O7}) are given here by
\cite{phoenix},

\begin{eqnarray}
V_{++} &=& \sqrt{2}\; \cos \left( \Omega \tau \sqrt{aa^{\dag}}\right) \nonumber \\
V_{+-} &=& -i\sqrt{2}\; \frac{\sin \left( \Omega \tau \sqrt{aa^{\dag}}\right) }{%
\sqrt{aa^{\dag}}}a\nonumber \\
V_{-+} &=& -V_{+-}^\dagger\nonumber \\
V_{--} &=& V_{++}^\dagger \label{O17}
\end{eqnarray}
 where $\tau$ is the interaction time (the time of flight of the atom through the
 resonator).\\
 \begin{figure}[h]
\includegraphics[scale=0.45]{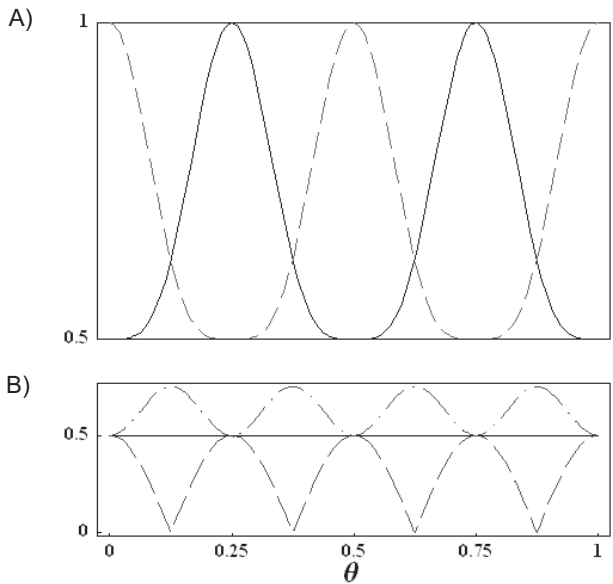} \caption{\label{fig2}(a) $P_Q$ (solid curve) and $P_M$ (dashed
curve). (b) Each one of the three terms in Eq. (\ref{O18}):
$|\mathcal{G}_Q-\mathcal{G}_M|$ (dashed), $\mathcal{G}$ (solid) and
$\mathcal{G}_Q+\mathcal{G}_M$ (dot-dashed). Both plots are shown as
functions of the vacuum Rabi phase $\theta$ for $\overline{n}_{o}=0$
and $s=0$. }
\end{figure}
\begin{figure}[h]
\includegraphics[scale=0.55]{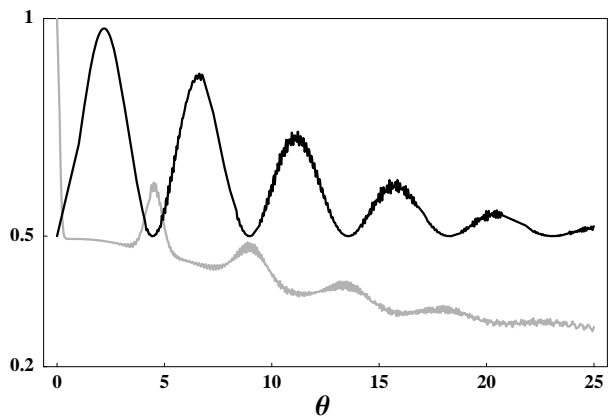} \caption{\label{fig3}$P_Q$ (black curve) and $P_M$ (grey curve) as
functions of the vacuum Rabi phase $\theta$ for
$\overline{n}_{o}=20$ and $s=0$. }
\end{figure}

Consider now the cavity field prepared in the vacuum state so that
the mean photon number is $\overline{n}_0=0$. The results for the
purity of both subsystems (Eqs. (\ref{O15}) and (\ref{O16})) are
shown in Fig. 2 versus the normalized Rabi phase
$\theta=\Omega\tau/{2\pi}$. Fig. 2(a) displays the dynamical process
of purity swapping \cite{sudarshan}. Here, the Bloch vector of the
quanton oscillates in lenght in counterphase with the purity of the
cavity field. As seen in the plot, both systems interchange purity
periodically, with a period $T=\pi/\Omega$. This interchange is
bounded by an Araki-Lieb type inequality as can be seen in Fig. 2(b)
where the inequality
 \begin{equation}
|\mathcal{G}_Q - \mathcal{G}_M|\leq \mathcal{G}\leq \mathcal{G}_Q +
\mathcal{G}_M \label{O18}
\end{equation}
is satisfied at all times. Moreover, we have numerically confirmed
that Eq. (\ref{O18}) is satisfied for a dense grid of values of
$(s,\,\overline{n}_0,\,\theta)$ \cite{footnote5}. This result
supports the conjecture given in Eq. (\ref{O13.7}). Note in Fig.
2(b) that the points of maximal purity interchange makes Eq.
(\ref{O18}) an equality. The oscillations shown in Fig. 2(a) are
similar to those found in \cite{sudarshan} for a pair
of qubits coupled by a nonlocal interaction.\\
\begin{figure}[h]
\includegraphics[scale=0.48]{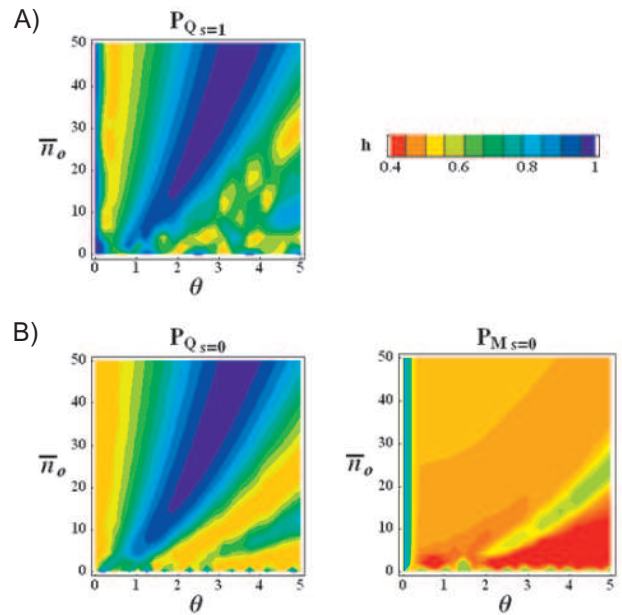} \caption{\label{fig4}Contour plots of $P_Q$ and $P_M$ as functions
of the cavity field's intensity $\overline{n}_{0}$ and the vacuum
Rabi phase $\theta$. Plots are shown for two different initial
preparations of the quanton's state. (a) The plot on first line
belongs to pure state preparation ($s=1$). (b) The remaining two
plots belong to a totally mixed state preparation ($s=0$). }
\end{figure}
But, what happens when we inject to the cavity field more and more
photons? In a typical experimental situation, the cavity field is
prepared in a coherent state with a large $\overline{n}_{0}$
\cite{raimond}. We find that purity swapping is still obtained. The
oscillations become damped and more irregular, since the dynamics
mixes different phases stemming from different photon manifolds. The
$P_M$ envelope decreases with $\theta$, indicating that more photon
manifolds get entangled as the number of Rabi floppings increases.
This is shown in Fig. 3 where $P_{Q, M}$ are plotted for $s=0$ and
$\overline{n}_0=20$. What we see here are manifestations of the
collapses and revivals of the JCM \cite{Gea90}. The first plateau in
$P_M$ corresponds to the collapse region ($\mathcal{P}\rightarrow
0$). In these zones, $P_Q$ follows the behavior of the visibility
$\mathcal{V}$. We show this explicitely in Fig. 4, where $P_{Q,M}$
is plotted for different preparations of $\overline{n}_0$ and
$\theta$.

On one hand, Fig. 4(a) shows results for initial pure state
preparation of the quanton ($s=1$). Here $P_Q=P_M$. We can
understand this effect in terms the Araki-Lieb inequality given in
Eq. (\ref{O18}). After Eq. (\ref{O13.3}) we have
$\mathcal{G}^{(0)}=0$. The purity is conserved under unitary $U$
global evolution, so $\mathcal{G}^{(f)}=0$. According to Eq.
(\ref{O18}), $\mathcal{G}_Q=\mathcal{G}_M$ at all times. Apart from
Eq. (\ref{O18}), this property can also be derived from the general
properties of $EM$. In fact, not only the entropy of entanglement
but all $EM$ for pure states are symmetric under the exchange of
parties \cite{vidal2000}. Another main feature of Fig. 4 can be
easily related to the properties of $EM$. In fact, for every $EM$
measure \cite{vidal2000}
\begin{equation}
EM(\rho)\geq0. \label{O18.001}
\end{equation}
For a separable state $\rho_{QM}$, $EM(\rho_{QM})=0$. This is what
occurs at the $P_Q\rightarrow 1$ zone of Fig. 4(a). The dynamics
decouples $\rho_Q$ and $\rho_M$ asymptotically in the recreation
zone defined by the relation
\begin{equation}
\theta_R=\sqrt{\overline{n}_0}. \label{O18.01}
\end{equation}
Here Eq. (\ref{O13.3}) is satisfied and reduces to
$\mathcal{G}_{Q,M}=0$, so $P_Q=P_M\rightarrow1$. The linear entropy
properties as an $EM$ accounts for the recreation of state vector
phenomena for $s=1$ found by Gea-Banacloche \cite{Gea90}.

On the other hand, new results are shown in Fig. 4(b) for the case
of a initial totally unpolarized quanton state (s=0). Remarkably, we
obtain here as well an asymptotically recreation of the state vector
$P_Q\rightarrow1$ in the first recreation zone. This is apparent
comparing the plots for ${P_Q}_{s=1}$ and ${P_Q}_{s=0}$ in Fig. 4(a)
and Fig. 4(b). Here, it can be seen that many revival zones wash out
in the lower plot. However, the recreation of the state vector in
the first zone $\theta_R$ is robust against degradation of the
initial purity of the quanton, given by $s$ (i.e.,
$P_Q^{(0)}=\frac{1}{2}(1+s^2)$). This is one of the main results of
the paper.

This results contrast the unitary WWM case, where $P_Q^{(0)}=0$
implies $P_Q^{(f)}=0$ at all times (see Eq. (\ref{O6.2})). Contrary
to this, we obtain here that the JCM interaction can result in a
significant increase of the visibility of a totally unpolarized
quanton. This can be seen in the right plot of Fig. 5(b), where the
visibility for $s=0$ is explicitly plotted in the same fashion as
Fig. 4. The blue zone gives a wide region for experimentalists
willing to obtain perfect interference patterns starting from
totally unpolarized sources. We can understand this phenomena by
recalling again Eq. (\ref{O18}). According to Eq. (\ref{O13.3}),
$\mathcal{G}_{Q}^{(0)}=\mathcal{G}^{(0)}>0$, since we start now with
a mixed quanton's state. Eq. (\ref{O18}) allows a net transfer of
entropy build up from the quanton to the cavity field. $P_Q\simeq1$
can arise at the expenses of maximally increasing the entropy of the
cavity field. This can be seen comparing $P_Q$ and $P_M$ in Fig.
4(b). The quanton gets pure $P^{(0)}_Q=1/2\longrightarrow
P^{(f)}_Q=1$ at the expense of a reciprocal increase of the entropy
of the interacting system $P^{(0)}_M=1\longrightarrow
P^{(f)}_M=1/2$. This maximal purity swapping is consistent with the
Araki-Lieb bounds of Eq. (\ref{O18}). In fact, they are not only
consistent but demanded by it. In order to see this, let us insert
Eqs. (\ref{O0.4}) and (\ref{O13.31}) into Eq. (\ref{O18}) for $s=0$.
We have
\begin{equation}
|P_M-P_Q|\leq\frac{1}{2}\leq 2-P_M-P_Q. \label{O18.012}
\end{equation}
Eq. (\ref{O18.012}) sets up the bounds of purity exchange between
the systems. This bounds can be observed in the results given in
Figs. (2-4). These are strong bounds and provide useful information.
For instance as $P_Q\rightarrow 1$, it is easy to show that Eq.
(\ref{O18.012}) demands $P_M\rightarrow 1/2$.

Both subsystems decouple in the recreation zone asymptotically with
$\overline{n}_0$ for all values of $s$. The mutual information
\begin{equation}
\mathcal{I}=\mathcal{G}_Q+\mathcal{G}_M-\mathcal{G}_Q
\mathcal{G}_M-\mathcal{G} \label{O18.02}
\end{equation}
is plotted in Fig. 5. It can be seen that for $s=0$ less
entanglement and wider zones of decoupling are obtained in
comparison to $s=1$. It can also be noticed in Fig. 5(b) that even
if we start from a totally mixed quanton state, an appreciable
amount of entanglement can build up at long times for low values of
$\overline{n}_0$.
\\
\begin{figure}[h]
\includegraphics[scale=.34]{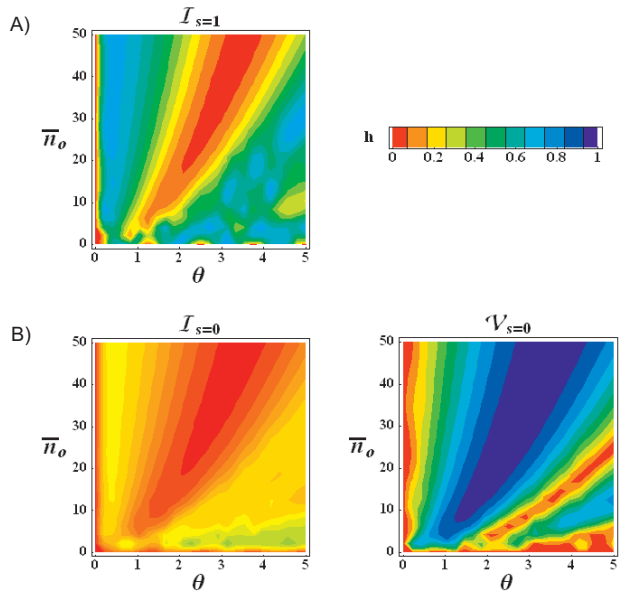} \caption{\label{Fig5} Mutual information of the combined system
and fringe visibility as functions of the cavity field's intensity
$\overline{n}_{o}$ and the vacuum Rabi phase $\theta$. Plots are
shown for two different initial preparations of the quanton's state:
(a) pure and (b) totally unpolarized.}
\end{figure}
Finally, we connect our results with the robustness of the Gea
attractor. Julio Gea-Banacloche \cite{Gea90} studied the $s=1$ case.
At the beginning of its evolution, the quanton becomes rapidly
unpolarized (collapse region), but right after the quanton evolves
to the form of the pure state attractor
\begin{equation}
\label{O18.03}
\left|\Psi\right\rangle_Q^{attr}=\frac{1}{\sqrt{2}}(\left|e\right\rangle+\,i\,e^{i\alpha}\left|g\right\rangle),
\end{equation}
where $\alpha$ is the phase of the cavity field. The above attractor
state arises at the half of the revival time leading to the
recreation of the state vector. As was demonstrated in \cite{Gea90},
the state of any initial totally polarized atom ($s=1$ case) will evolve to the
attractor state, regardless of any other atomic initial conditions.

Now, we show that the Gea-Banacloche attractor state is reached also
for any initial purity of the state. Towards this goal we calculate
the state just after the beam splitter. We can undo the action of
the beam merger by taking the transformation on Eqs. (\ref{O20})
\begin{equation}
\label{O21}
\sigma_x\rightarrow\sigma_z,\;\;\sigma_z\rightarrow-\sigma_x.
\end{equation}

With these transformations, taking $\phi=0$ and taking the trace of
the resulting total state over the WWM's degree of freedom we obtain
the quanton's state just after the beam splitter

\begin{equation}
\label{O22} \rho_Q^{BS}=\frac{1}{2}\left(1-i\sigma_x \Im m
\left[\mathcal{C}\right]+\sigma_y\Re e
\left[\mathcal{C}\right]+\sigma_z\mathcal{P}\right),
\end{equation}
where $\mathcal{P}$ and $\mathcal{C}$ were already defined in Eqs.
(\ref{O2}) and (\ref{O7.5}). Now we particularize the above state to
the recreation zone given by Eq. (\ref{O18.01}). As seen in Fig.
4(b), here $\mathcal{V}\rightarrow 1$. Thus $\mathcal{P}\rightarrow
0$, since $\mathcal{V}^2+\mathcal{P}^2\leq1$. Therefore, using Eq.
(\ref{O7.4}) it is easy to show that in the recreation zone Eq.
(\ref{O22}) tends to Eq. (\ref{O18.03}), once $\alpha$ is defined as
the phase of the complex contrast factor $\mathcal{C}$
\cite{footnote6}. The quanton state evolves to the pure state
Gea-Banacloche attractor, gaining purity at the expense of
increasing the entropy on the cavity field, which gets decoupled
from the quanton in the process. Note that in this calculation we
did not particularize at anytime for any initial quanton's state.
Thus, we have generalized the result from \cite{Gea90} and
demonstrated that the Gea-Banacloche attractor is robust against all
quanton's initial conditions, including degradation of the purity of
the quanton.

\section{Conclusions}
In conclusion, we have shown the existence of dynamical purity
swapping in the JCM. The dynamic itself purifies the qubit at the
expense of degrading the purity of the cavity field. Moreover, we
have shown that a qubit can exhibit dynamical purity swapping with a
generic quantum system, provided they coupled via a non-unitary
matrix elements interaction [in the sense of Eq. (\ref{O7})]. We
have been able to obtain such general necessary condition for purity
swapping thanks to an interferometric approach, allowing us to
connect purity degradation with which-way marking. Then we have
analyzed in detail the particular case of the JCM, since it
describes a large variety of systems. It also serves as a
cornerstone for experimental quantum information, communication and
computing. In fact, the observation of this phenomena is perfectly
attainable with current technology \cite{newHaroche2001}. We have shown that the Gea-Banacloche attractor is robust against
degradation of the initial purity of the quanton.
Any initially totally unpolarized qubit will evolve to the pure state Gea-Banacloche attractor after interacting with the cavity field the time required by Eq. (\ref{O18.01}). Thus, we can use the collapses and revivals phenomena of the JCM for dynamical purification of qubits.
Since the field's phase $\alpha$ in Eq. (\ref{O18.03}) is an externally controllable parameter, this phenomena can also be used for quantum preparation of pure superposition states starting from totally mixed states.
This demonstrates
in addition the possibility of a remarkable phenomena: the arising
of a perfect visibility interference pattern starting from a totally
unpolarized source of qubits.
Finally, we show that the Tsallis entropy $T_2$  is a useful entanglement monotone ($EM$) allowing one to relate entanglement with purity swapping. Many features of the phenomena have been shown to derive from the algebraic properties of $EM$.


%

\begin{acknowledgments}
This research was supported by a Return Program from the
Consejer\'{i}a de Educaci\'on y Ciencia de la Junta de Andaluc\'{i}a
in Spain.
\end{acknowledgments}

%
%


%
%
%


\begin{references}
\bibitem{footnote4} To our knowledge, this inequality has not yet
been proved.
\bibitem{footnote2} As noted in \cite{Englert96b}, there is
no need to consider more generic Bloch vector, since this just
amounts to a redefinition of the operators $V_{\pm\pm}$ in Eq.
(\ref{O7}).
\bibitem{footnote5} We have checked Eq. (\ref{O18}) over 1600 points
in the interval
$\overline{n}_0\,\epsilon\,(0,\,1000),\,\theta\,\epsilon\,(0,\,200\pi)$,
$s\,\epsilon\,(0,1)$.
\bibitem{footnote6}In fact, according to Eqs. (\ref{O7.8}) and
(\ref{O17}) $\alpha$ can be shown to be the phase of the cavity
field.
\end{references}
\end{document}